\documentclass[preprint,showpacs,showkeys,aps,prb]{revtex4}
\usepackage[dvips]{graphicx}
\begin{document}

\title{
Singly-Thermostated Ergodicity in Gibbs' Canonical Ensemble \\
and the 2016 Ian Snook Prize Award \\
}

\author{
William Graham Hoover and Carol Griswold Hoover               \\
Ruby Valley Research Institute                  \\
Highway Contract 60, Box 601                    \\
Ruby Valley, Nevada 89833                       \\
}

\date{\today}


\vspace{0.1cm}

\begin{abstract}
The 2016 Snook Prize has been awarded to Diego Tapias, Alessandro Bravetti, and
David Sanders for their paper ``Ergodicity of One-Dimensional Systems Coupled to
the Logistic Thermostat''.  They introduced a relatively-stiff hyperbolic tangent
thermostat force and successfully tested its ability to reproduce Gibbs' canonical
distribution for three one-dimensional problems, the harmonic oscillator, the
quartic oscillator, and the Mexican Hat potentials :
$$
\{ \ (q^2/2) \ ; \ (q^4/4) \ ; \ (q^4/4) - (q^2/2) \ \} \ .
$$
Their work constitutes an effective response to the 2016 Ian Snook Prize Award
goal, ``finding ergodic algorithms for Gibbs’ canonical ensemble using a single
thermostat''.  We confirm their work here and highlight an interesting feature
of the Mexican Hat problem when it is solved with an adaptive integrator.

\end{abstract}

\maketitle

\section{Nos\'e and Nos\'e-Hoover Canonical Dynamics Lack Ergodicity}
In 1984 Shuichi Nos\'e used ``time scaling''\cite{b1,b2} to relate his novel
Hamiltonian ${\cal H}$ to an extended version of Gibbs' canonical phase-space
distribution $f$ , proportional to $e^{-{\cal H}/kT}$ . Hoover's simpler
``Nos\'e-Hoover'' motion equations\cite{b3} dispensed with Hamiltonian
mechanics and time scaling, reducing the dimensionality of the extended phase
space by one. For the special case of a harmonic oscillator the Nos\'e-Hoover
motion equations and the corresponding modified Gibbs' distribution are :
$$
\{ \ \dot q = p\ ; \ \dot p = - q - \zeta p \ ;
\ \dot \zeta = [ \ (p^2/T) - 1 \ ]/\tau^2 \ \} \ \longrightarrow
$$
$$
f(q,p,\zeta) \propto e^{-(q^2/2T)} e^{-(p^2/2T)} e^{-(\zeta^2 \tau^2/2)} \ 
[ \ {\rm Nos\acute{e}-Hoover} \ ] \ .
$$
Here $q$ and $p$ are the oscillator coordinate and momentum.  $\zeta$ is a
``friction coefficient'', or ``control variable''. In all that follows we
choose the equilibrium temperature $T$ equal to unity to simplify notation.
The timescale of the thermal response to the imposed equilibrium temperature
is governed by the relaxation time $\tau$ . For simplicity we choose force
constants, masses, and Boltzmann's constant all equal to unity.

Hoover used the steady-state phase-space continuity equation :
$$
(\partial f/\partial t) = -\nabla \cdot (fv) = 0 \ ,
$$
to show that Gibbs' canonical distribution is consistent with the Nos\'e-Hoover
motion equations.  Here the phase-space flow velocity is
$v \equiv (\dot q,\dot p,\dot \zeta)$ .
Hoover's numerical work showed that only a portion of the three-dimensional
Gaussian distribution (typically just a two-dimensional torus) is generated.
That is, solutions of these three-dimensional motion equations are not ergodic.
Particular solutions fail to cover the entire $(q,p,\zeta)$ phase space.

\begin{figure}
\includegraphics[width=4.5in,angle=+90.0]{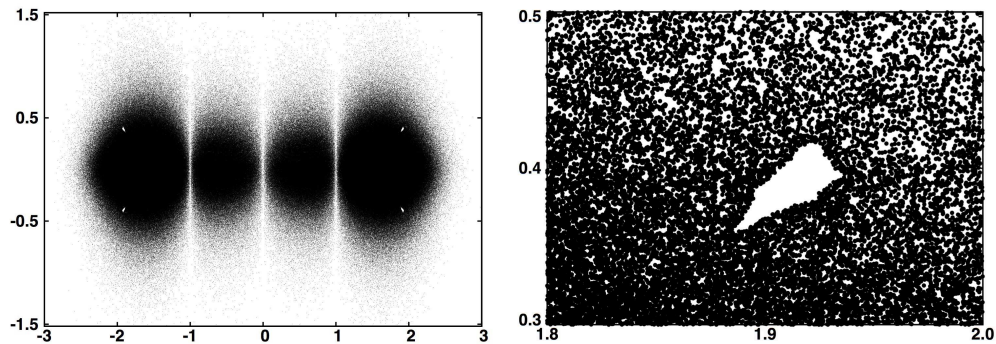}
\caption{
The $p=0$ Mexican Hat cross section for $\alpha = 6.5$ has four apparent holes,
one of which is shown in the closeup to the right.  Here, and also in Figures
2 and 3, the abscissa is $q$ and the ordinate is the friction coefficient $\zeta$ .
}
\end{figure}

\begin{figure}
\includegraphics[width=4.5in,angle=+90.0]{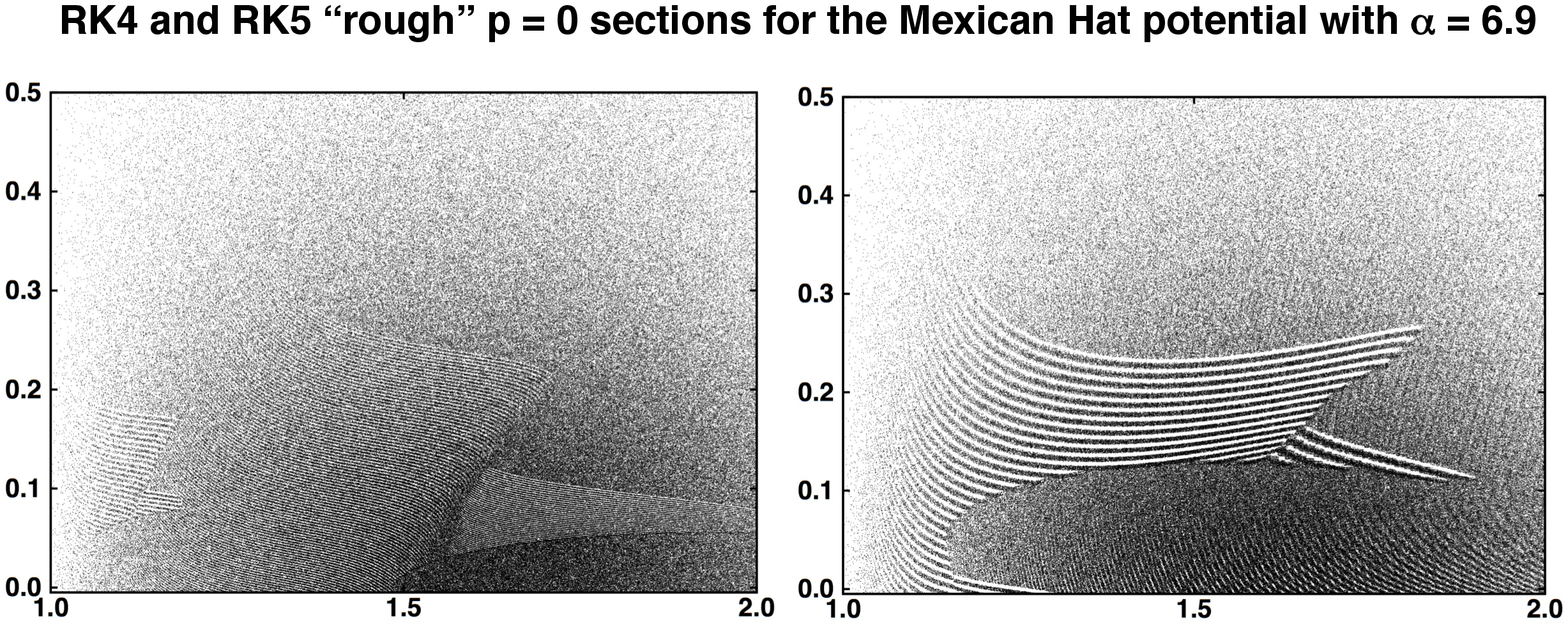}
\caption{
The $p=0$ Mexican Hat cross section closeups for $\alpha = 6.9$ with the rms
difference between solutions with timesteps of $dt$ and two steps of $(dt/2)$
constrained to lie in the range
$10^{-14} \ {\rm to} \ 10^{-12}$.
The fourth-order section is on the left and the fifth-order section is on the right. 
}
\end{figure}

\begin{figure}
\includegraphics[width=4.5in,angle=-90.0]{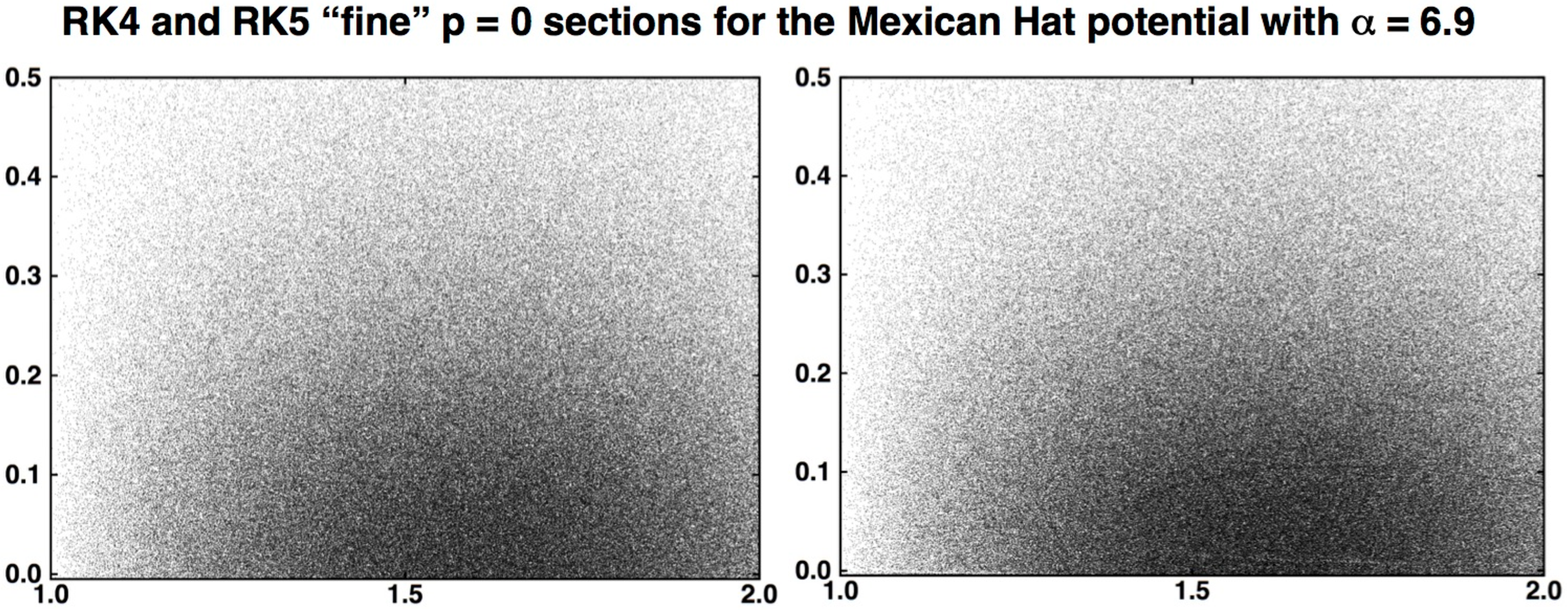}
\caption{
  The $p=0$ Mexican Hat cross section closeups for $\alpha = 6.9$ with the rms
difference between solutions with timesteps of $dt$ and two steps of $(dt/2)$
constrained to lie in the range
$10^{-17} \ {\rm to} \ 10^{-15}$.
The fourth-order section is on the left and the fifth-order section is on the right.
}
\end{figure}

Considerable numerical work, following the comprehensive analyses of Kusnezov,
Bulgac, and Bauer\cite{b4,b5}, suggested that using {\it two} thermostat variables
rather than {\it one} was the simplest route to oscillator ergodicity. Including
another thermostat variable requires a {\it four}-dimensional $(q,p,\zeta,\xi)$
phase space.  A successful example\cite{b6}, ergodic in $(q,p,\zeta,\xi)$ space,
 controlled two velocity moments, $\langle \ p^2 \ \rangle$ and
$\langle \ p^4 \ \rangle$ rather than just one :
$$
\{ \ \dot q = p\ ; \ \dot p = - q - \zeta p - \xi p^3\ ;
\ \dot \zeta = p^2 - 1  \ ; \ \dot \xi = p^4 - 3p^2 \ \} \ \longrightarrow
$$
$$
f(q,p,\zeta) \propto e^{-(q^2/2)} e^{-(p^2/2)} e^{-(\zeta^2/2)} e^{-(\xi^2/2)} \
[ \ {\rm Hoover-Holian } \ ] \ .
$$

In 2015 a {\it single}-thermostat approach\cite{b7} with simultaneous weak control
of $\langle \ p^2 \ \rangle$ and $\langle \ p^4 \ \rangle$ was found to generate
Gibbs' entire distribution for the harmonic oscillator :
$$
\{ \ \dot q = p \ ; \ \dot p = -q -  \zeta (0.05p + 0.32p^3) \ ; \ \dot \zeta =
0.05( p^2 - 1) + 0.32( p^4 - 3 p^2) \ \} \ . 
$$
Straightforward generalizations of this single-thermostat approach failed to
thermostat the quartic and Mexican Hat potentials, leading to the posing of the 2016
Snook Prize problem solved by Tapias, Bravetti, and Sanders\cite{b8}.

\section{Tapias, Bravetti, and Sanders' ``Logistic'' Thermostat}
The Logistic Map and the Logistic Flow are two simple models for chaotic 
behavior :
$$
q_{n+1} = cq_n(1 - q_n) {\rm \ and \ }  \dot q = q(1 - q) \ .
$$
A solution of the logistic flow equation is 
$$
\dot q = \frac{1}{ [ \ e^{+t/2} + e^{-t/2} \ ]^2} \longleftrightarrow 
q = \frac{e^{+t/2}}{[ \ e^{+t/2} + e^{-t/2} \ ]} \longleftrightarrow 2q = 1 + \tanh(+t/2) \ .
$$
With these logistic equations in mind Tapias, Bravetti, and Sanders\cite{b8} suggested
a hyperbolic tangent form for the thermostat variable, and showed, with a variety of
numerical techniques, convincing evidence for the ergodicity of their ``Logistic Thermostat''
 motion equations for the quartic and Mexican Hat potentials as well as the simpler
harmonic oscillator problem.

In the most challenging case, the Mexican Hat potential, the ergodic set of motion
equations found by Tapias, Bravetti, and Sanders was feasible to solve, but relatively
stiff :
$$
\{ \ \dot q = p \ ; \ \dot p = q - q^3 - 50p\tanh(25\zeta)
 \ ; \ \dot \zeta = p^2 - 1 \ \} \ .
$$
In replicating their work we also characterized solutions of a slight variant :
$$
\{ \ \dot q = p \ ; \ \dot p = q - q^3 - \alpha p\tanh(\alpha \zeta)
 \ ; \ \dot \zeta = p^2 - 1 \ \} \ ,
$$
where values of the parameter $\alpha$ in the neighborhood of seven lead to apparent
ergodic behavior in $(q,p,\zeta)$ space.

One of the simplest and most useful tests for ergodicity in three dimensions is the
lack of holes in the two-dimensional cross-sections (as opposed to projections) of
the three-dimensional flow. For stiff equations it is convenient to use ``adaptive''
integrations of the motion equations where the timestep varies to maintain the
accuracy of the integrator\cite{b9}.

In our own  numerical work we integrated for a time of 10,000,000 using timesteps
which maintained the rms difference between a fourth-order or fifth-order Runge-Kutta
step of $dt$ and two such  steps with $(dt/2)$ to lie within a band varying from
$$
10^{-12} > \sqrt{ \delta q^2 + \delta p^2 + \delta \zeta^2 } > 10^{-14} \ {\rm to} \
10^{-15} > \sqrt{ \delta q^2 + \delta p^2 + \delta \zeta^2 } > 10^{-17} \ .
$$
We generated about 3,000,000 $\{ \ q,0,\zeta \ \}$ double-precision cross-section
points in laptop runs taking about an hour each. Typical timesteps were in the range
from 0.0001 to 0.001 .

Figure 1 shows portions of the $p=0$ cross section with $\alpha = 6.5$ which has evident
holes at $(q=\pm 1.92,\zeta=\pm 0.39)$ .  The holes disappear if $\alpha$ is increased to
6.9. But a look at the $(q,0,\zeta)$ section with an error band of $10^{-13\pm 1}$ reveals
not only ``normal'' (irregularly-dotted) regions but also a few {\it striped} regions.
In Figure 2 we see that the stripes using RK4 differ from those using RK5 showing
that the stripes are artefacts.  Tightening the error band to $10^{-15\pm 1}$ confirms
this diagnosis, as shown in Figure 3.  The interesting structure of these striped regions
is a thoroughly unexpected fringe benefit of the new logistic thermostat.

We thank Drs Tapias, Bravetti, and Sanders for their stimulating prize-winning work.


\begin{thebibliography}{99}

\bibitem{b1}  S. Nos\'e, ``A Unified Formulation of the Constant Temperature Molecular
              Dynamics Methods'', Journal of Chemical Physics {\bf 81}, 511-519 (1984).

\bibitem{b2}  S. Nos\'e, ``Constant Temperature Molecular Dynamics Methods'', Progress
              in Theoretical Physics Supplement {\bf 103}, 1-46 (1991).

\bibitem{b3}  Wm. G. Hoover, ``Canonical Dynamics: Equilibrium Phase-Space Distributions'',
              Physical Review A {\bf 31}, 1695-1697 (1985).

\bibitem{b4}  D. Kusnezov, A. Bulgac, and W. Bauer, ``Canonical Ensembles from Chaos'',
              Annals of Physics {\bf 204}, 155-185 (1990).

\bibitem{b5}  D. Kusnezov and A. Bulgac, ``Canonical Ensembles from Chaos: Constrained
              Dynamical Systems'', Annals of Physics {\bf 214}, 180-218 (1992).

\bibitem{b6}  Wm. G. Hoover and B. L. Holian, ``Kinetic Moments Method for the Canonical
              Ensemble Distribution'', Physics Letters A {\bf 211}, 253-257 (1996).

\bibitem{b7} Wm. G Hoover and C. G. Hoover, ``Singly-Thermostated Ergodicity in Gibbs'
             Canonical Ensemble and the 2016 Ian Snook Prize'', Computational Methods
             in Science and Technology {\bf 22}, 127-131 (2016) .

\bibitem{b8}  D. Tapias, A. Bravetti, and D. P. Sanders, ``Ergodicity of One-Dimensional
              Systems Coupled to the Logistic Thermostat'', Computational Methods in
              Science and Technology (in press, 2017) = arXiv 1611.05090 .

\bibitem{b9} W. G. Hoover and C. G. Hoover, ``Comparison of Very Smooth Cell-Model
             Trajectories Using Five Symplectic and Two Runge-Kutta Integrators'',
             Computational Methods in Science and Technology {\bf 21}, 109-116 (2015).

\end{thebibliography}
\end{document}